\documentclass[twoside]{LCWS11}
\usepackage[latin1]{inputenc}
\usepackage[dvips]{graphicx,epsfig,color}
\usepackage{wrapfig,rotating}
\usepackage{amssymb,amsmath,array}
\usepackage{subfig}
\usepackage{cite}
\begin{document}

\title{
%%%%   Paper title goes here  %%%%%%%%%%%%%%
The pure $B-L$ model and future linear colliders: \\ the Higgs sector} %% 
%***********************************************************************
% AUTHORS INFORMATION AREA
%***********************************************************************
\author{Lorenzo Basso$^1$, Stefano Moretti$^2$ and Giovanni Marco Pruna$^3$\thanks{Speaker: G.~M.~Pruna.}
% Optional short acknowledgment: remove next line if non-needed
% DO NOT MODIFY THE FOLLOWING '\vspace' ARGUMENT
\vspace{.3cm}\\
% Addresses and institutions (remove "1- " in case of a single institution)
1- Albert-Ludwigs-Universit\"at - Fakult\"at f\"ur Mathematik und Physik \\
D - 79104 Freiburg i.Br. - Germany
%% Remove the next three lines in case of a single institution
\vspace{.1cm}\\
2- University of Southampton - School of Physics \& Astronomy \\
Highfield, Southampton, SO17 1BJ - United Kingdom
%% Remove the next three lines in case of a single institution
\vspace{.2cm}\\
3- TU Dresden - Institut f\"ur Kern- und Teilchenphysik \\
Zellescher Weg 19, 01069 Dresden - Germany\\
}
%%***********************************************************************
% END OF AUTHORS INFORMATION AREA
%***********************************************************************

\maketitle

\begin{abstract}
We summarise the phenomenology of the Higgs sector of the 
minimal $B-L$ extension of the Standard Model at an $e^+e^-$ Linear 
Collider. Within such a 
scenario, we show that (in comparison with the Large Hadron Collider) 
several novel production and decay channels involving the two physical Higgs states could experimentally be accessed at such a machine. In particular, we present the scope of the $Z'$ strahlung process for single and double Higgs production, the only suitable mechanism for accessing an almost decoupled heavy scalar state.
\end{abstract}

\section{The $B-L$ model}

The ``pure'' $B-L$ (``baryon'' minus ``lepton'') model
is a peculiar choice from the the subset of the minimal Standard Model ($SM$) extensions obeying the $SU(3)_C\times
SU(2)_L\times U(1)_Y\times U(1)_{B-L}$ gauge symmetry. It is defined by precluding the mixing between the two $U(1)$ groups \cite{Basso:2008iv}.

In this report we will analyse some peculiar processes of the minimal
$B-L$ extension of the Standard Model at future $e^+e^-$ Linear
Colliders (LCs). For all practical purposes, we generally refer to an International Linear Collider ($ILC$) for a sub-TeV/TeV prototype and to a Compact LInear Collider ($CLIC$) for a TeV/multi-TeV design.

Firstly we will analyse the cross section and luminosity impact on
the significance of the Higgs-strahlung (off $Z'$) process, with
emphasis on both single and double Higgs production. Finally, we
will consider the case in which an Higgs state is produced in association with heavy neutrinos from direct $Z'$ production.

All these channels are based on the direct production of a $Z'_{B-L}$ boson (which is generally dominantly coupled to leptons \cite{Basso:2008iv}), hence a lepton collider is the most suitable environment for probing the considered model.

Finally, we remark that all the choices of parameters used in the following analysis are compatible with both experimental and theoretical constraints (see Reference~\cite{par:spa}). 

\section{Higgs-strahlung off $Z'$: single Higgs production}

Among the production mechanisms of the minimal $B-L$ model, a 
non-$SM$-like one for producing a single Higgs boson final state (either the light or heavy one) is Higgs-strahlung
off a $Z'$ boson, i.e., $e^+ e^- \to Z'\, h_{1,2}$.

\begin{figure}[!ht]
  \subfloat[]{ 
  \label{H1Zp_1TeV}
  \includegraphics[angle=0,width=0.45\textwidth ]{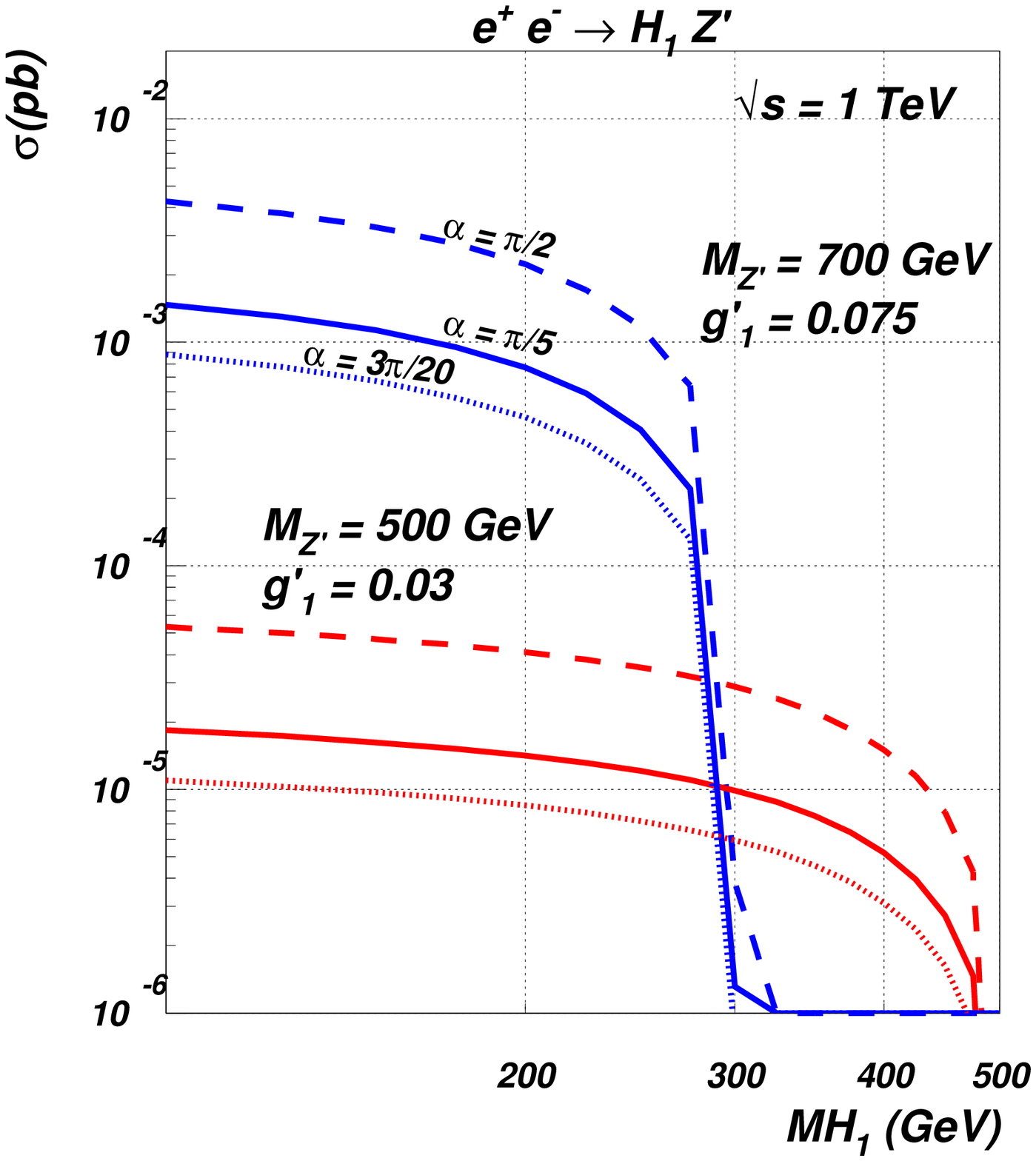}}
  \subfloat[]{
  \label{H1Zp_3TeV}
  \includegraphics[angle=0,width=0.45\textwidth ]{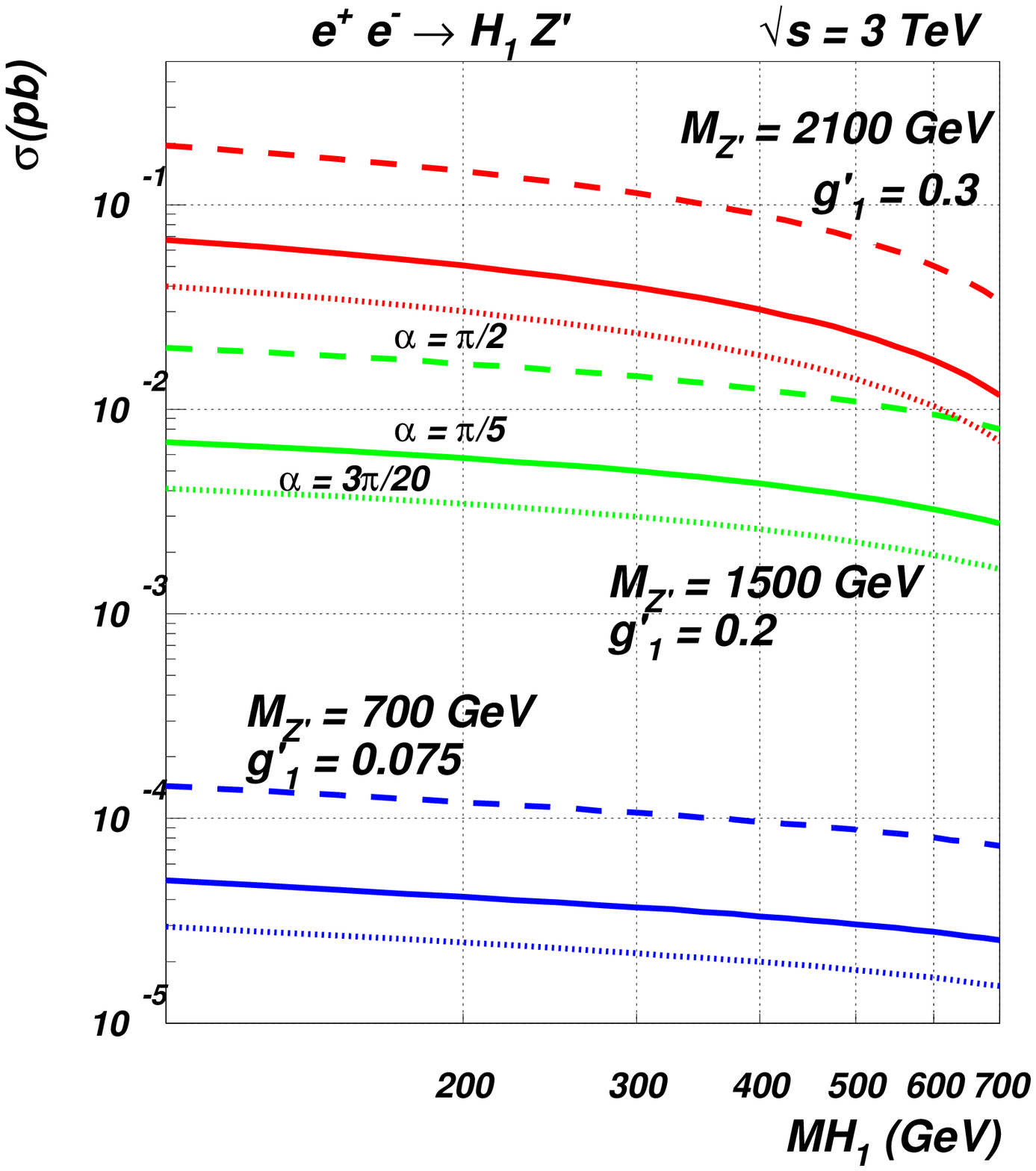}}
  \caption{\it Cross sections for the process
    $e^+e^-\rightarrow Z^{'\ast} \rightarrow H_{1} Z'$ at $\sqrt{s}=1$ (\ref{H1Zp_1TeV}) and $\sqrt{s}=3$ (\ref{H1Zp_3TeV})
    TeV. \label{H1Zp}}
\end{figure}

As shown in Figure~\ref{H1Zp_1TeV}(\ref{H1Zp_3TeV}) a light Higgs boson can be
produced in association with a $Z'$ boson of $1.5$($2.1$) TeV mass with cross
sections of $\mathcal{O}(10)$($\mathcal{O}(100)$) fb. Only the light Higgs boson has been
considered, being the case of a heavy Higgs boson with same mass just
the symmetric one under $\alpha \to \pi/2 - \alpha$.

It is important to remark that for $\alpha \to 0$ (decoupling limit), \emph{this is the only heavy Higgs
production mechanism allowed}. Moreover, the Higgs-strahlung off
$Z'$ mechanism is not suitable for the LHC \cite{Basso:2010yz}, making
a multi-TeV linear collider the ultimate chance for its
discovery. This mechanism is absent in scalar extensions of the $SM$
in which the gauge content is not changed. 

\begin{figure}[!ht]
  \subfloat[]{ 
  \label{a15}
  \includegraphics[angle=0,width=0.45\textwidth ]{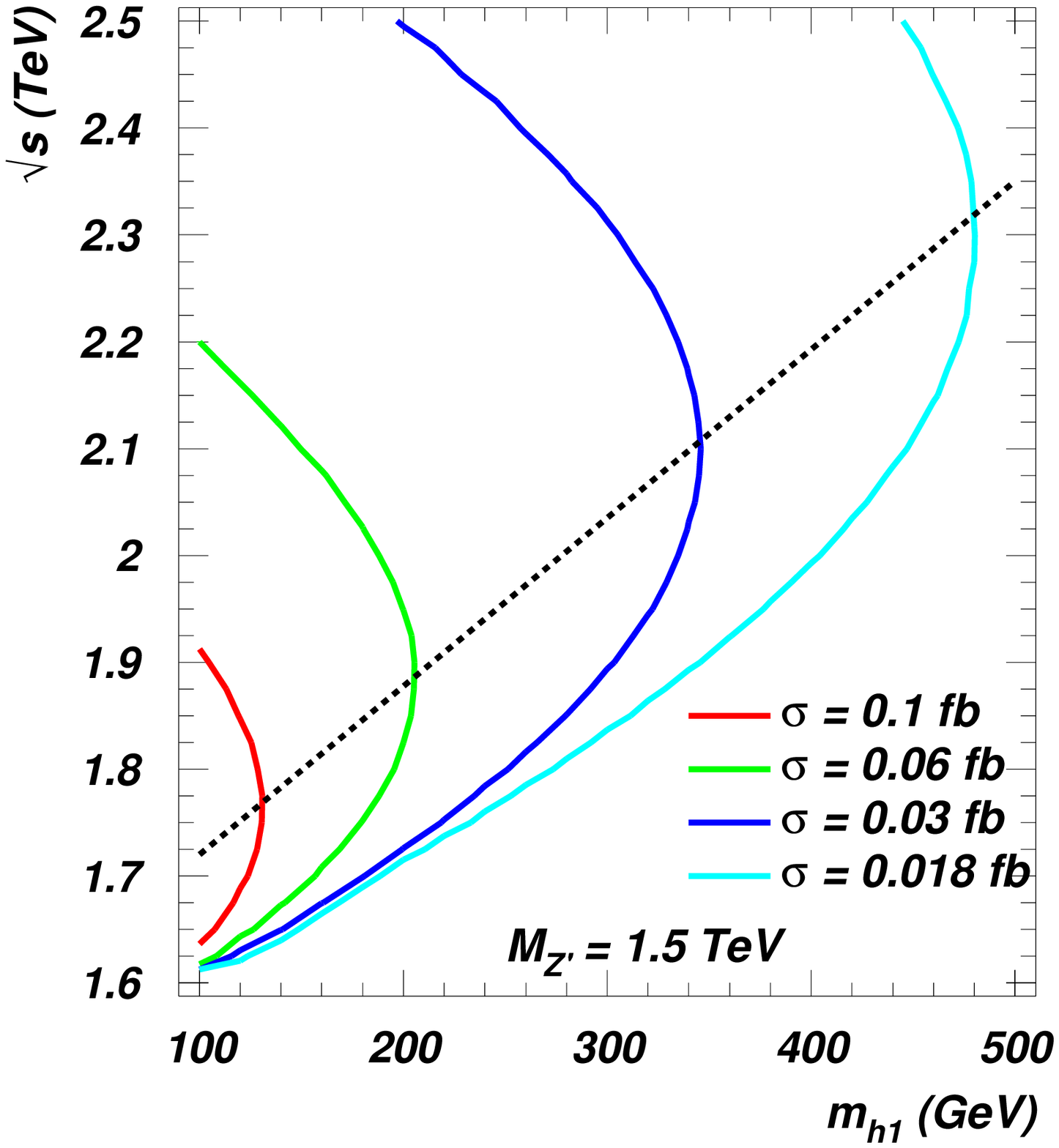}}
  \subfloat[]{
  \label{a21}
  \includegraphics[angle=0,width=0.45\textwidth ]{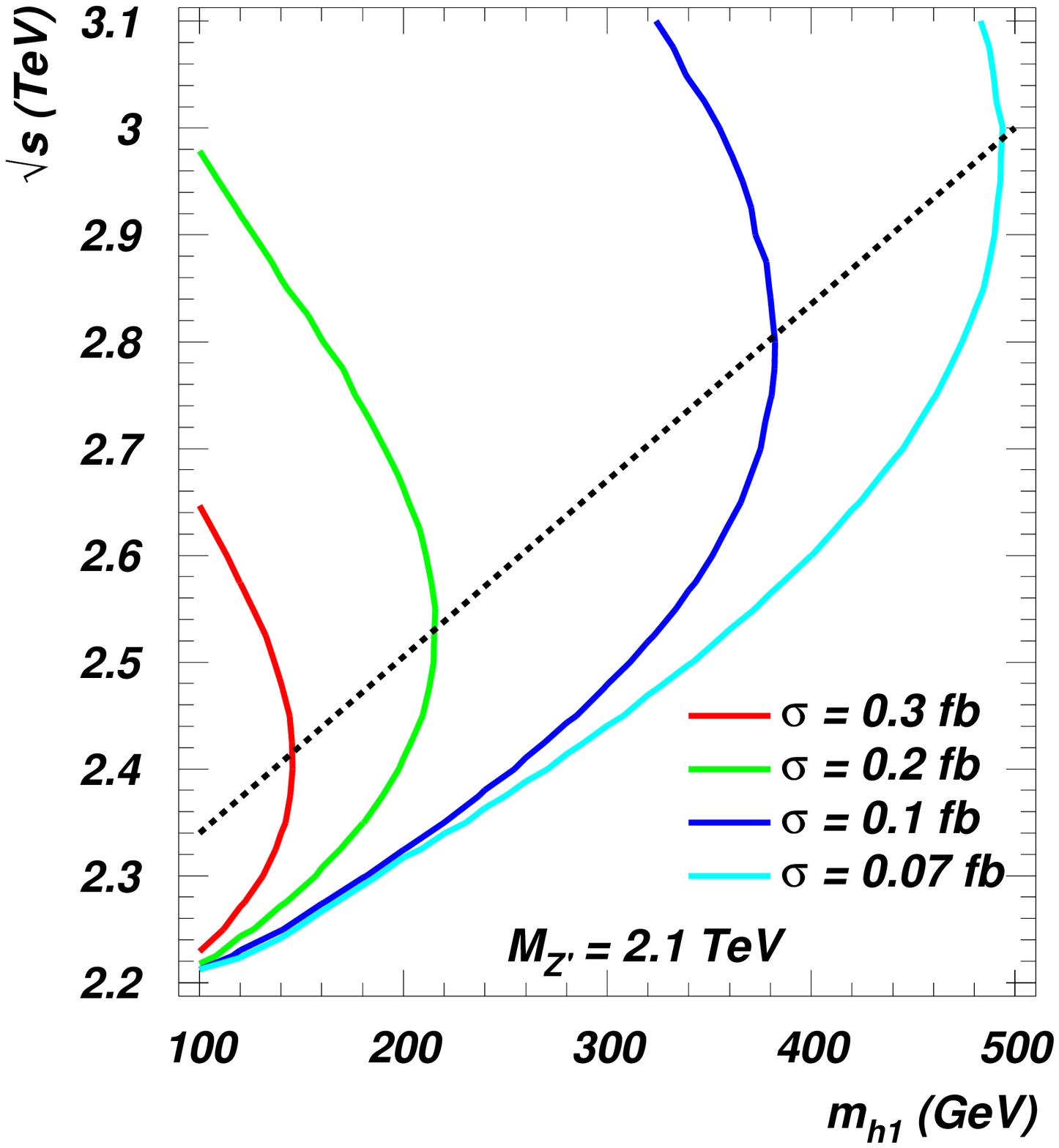}}
  \caption{\it ISR effect on the Higgs-strahlung off $Z'$ for
    $M_{Z'}=1.5$ TeV (\ref{a15}) and $M_{Z'}=2.1$ TeV (\ref{a21}).
    The dashed lines correspond to the $\sqrt{s}$
    for which the cross section per fixed Higgs mass is maximised,
    according to eq.~(\ref{Zp_ISR_dep}).\label{a}}
\end{figure}

With $Z'$ boson mass assumed to be known 
(see, for example, References~\cite{Basso:2009hf,Basso:2010pe}), the discovery potential
for this channel is here presented. We first study the effect of 
Initial State Radiation (ISR)
on the cross sections for the strahlung of a light Higgs state. 
Figure~\ref{a} clearly shows a linear dependence of
$\sqrt{s_{MAX}}$, the Centre-of-Mass (CM) energy that maximises the cross section, 
as a function of the Higgs mass only. Interpolating, we find the
useful relation:
\begin{equation}\label{Zp_ISR_dep}
\frac{\sqrt{s_{MAX}}}{TeV}\approx \frac{M_{Z'}}{TeV}+0.1+1.5
\frac{m_H}{TeV} \qquad (H=h_1, h_2).
\end{equation}
Hence, per fixed Higgs and $Z'$ boson masses, the discovery potential can be
maximised by fixing the CM energy to $\sqrt{s_{MAX}}$. 

As for the integrated luminosity required to start probing the values
of $\alpha$ in the $B-L$ model, benchmark scenarios use the usual
$h_1\rightarrow b\bar{b}$ ($m_{h_1}=120$ GeV) and $h_1\rightarrow
W^+W^-$ ($m_{h_1}=200$ GeV) decays. In analogy with the study of
Reference~\cite{Basso:2009hf}, the decay mode $Z'\rightarrow \mu^+\mu^-$ is
considered as the most suitable.
Regarding the background, the relevant one is found to be
$Z'Z/\gamma$ ($Z'\rightarrow \mu^+ \mu^-$ and $Z\rightarrow b\bar{b}$)
and $Z'W^+W^-$ (where, again, the muons come exclusively from the $Z'$
boson). The pure $EW$ background is two orders of magnitude below the
latter two, hence it is neglected here.

For both signal and background, we have assumed standard
acceptance cuts (for muons and quarks) at a LC~\cite{Assmann:2000hg},
and a window on the invariant mass $m(b,\bar{b}/W^+,W^-)=m_{h_1}\pm
20$ GeV to roughly emulate the detector resolution. For the
muons, we require $m(\mu^+,\mu^-)\in M_{Z'}\pm
1.5\Gamma_{Z'}$, always wider than the di-muon resolution for the
values of the gauge coupling here
considered~\cite{Basso:2008iv,Assmann:2000hg}. Finally, $b$-quark
tagging efficiency has been assumed to be $62\%$ according to
Reference~\cite{Abe:2001pea}. The $W$-boson reconstruction efficiency has
been set to $100\%$ as a reasonable approximation.

\begin{figure}[!ht] 
  \subfloat[]{
  \label{lumi120_2100}
  \includegraphics[angle=0,width=0.45\textwidth ]{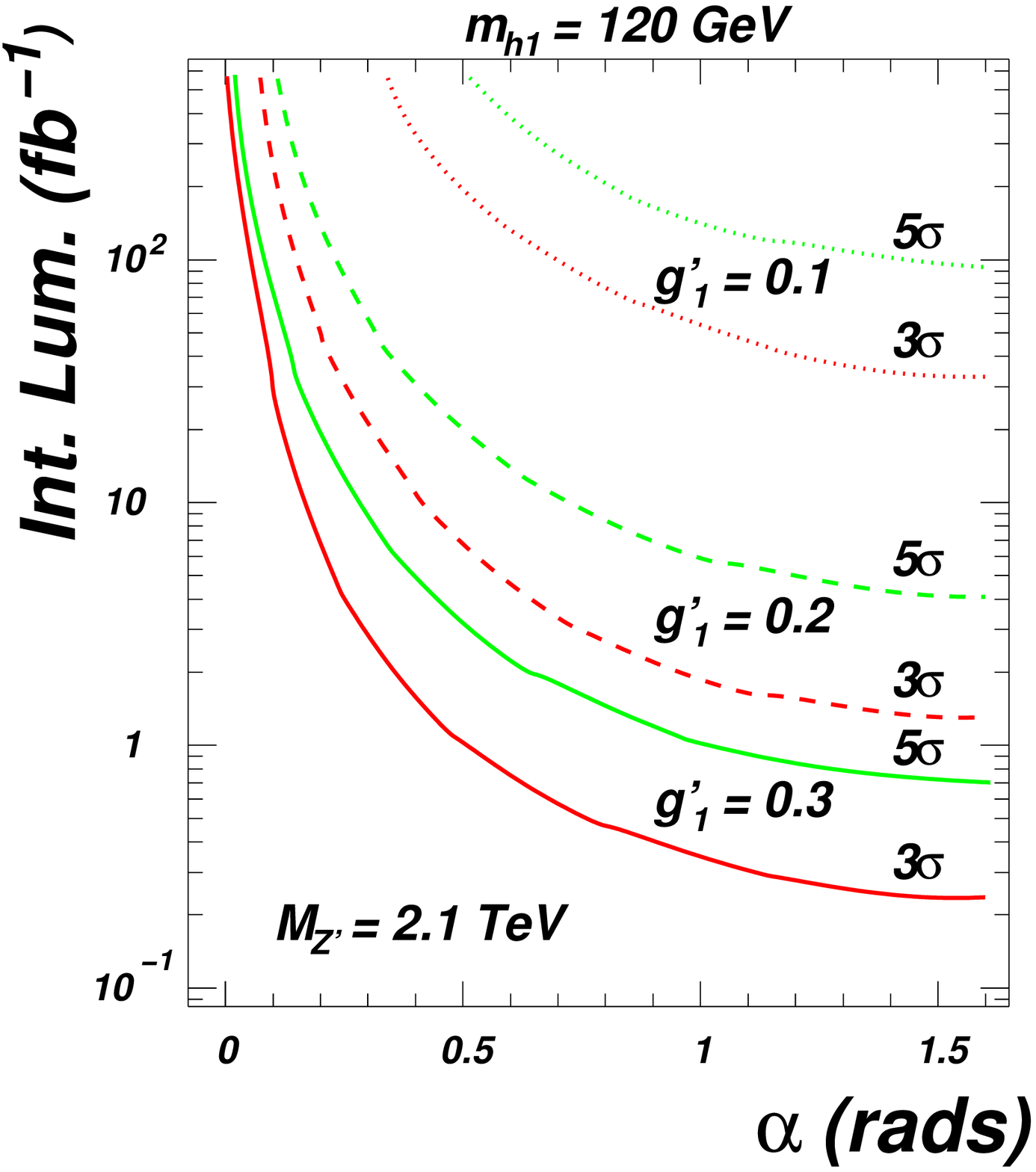}}
  \subfloat[]{
  \label{lumi200_2100}
  \includegraphics[angle=0,width=0.45\textwidth ]{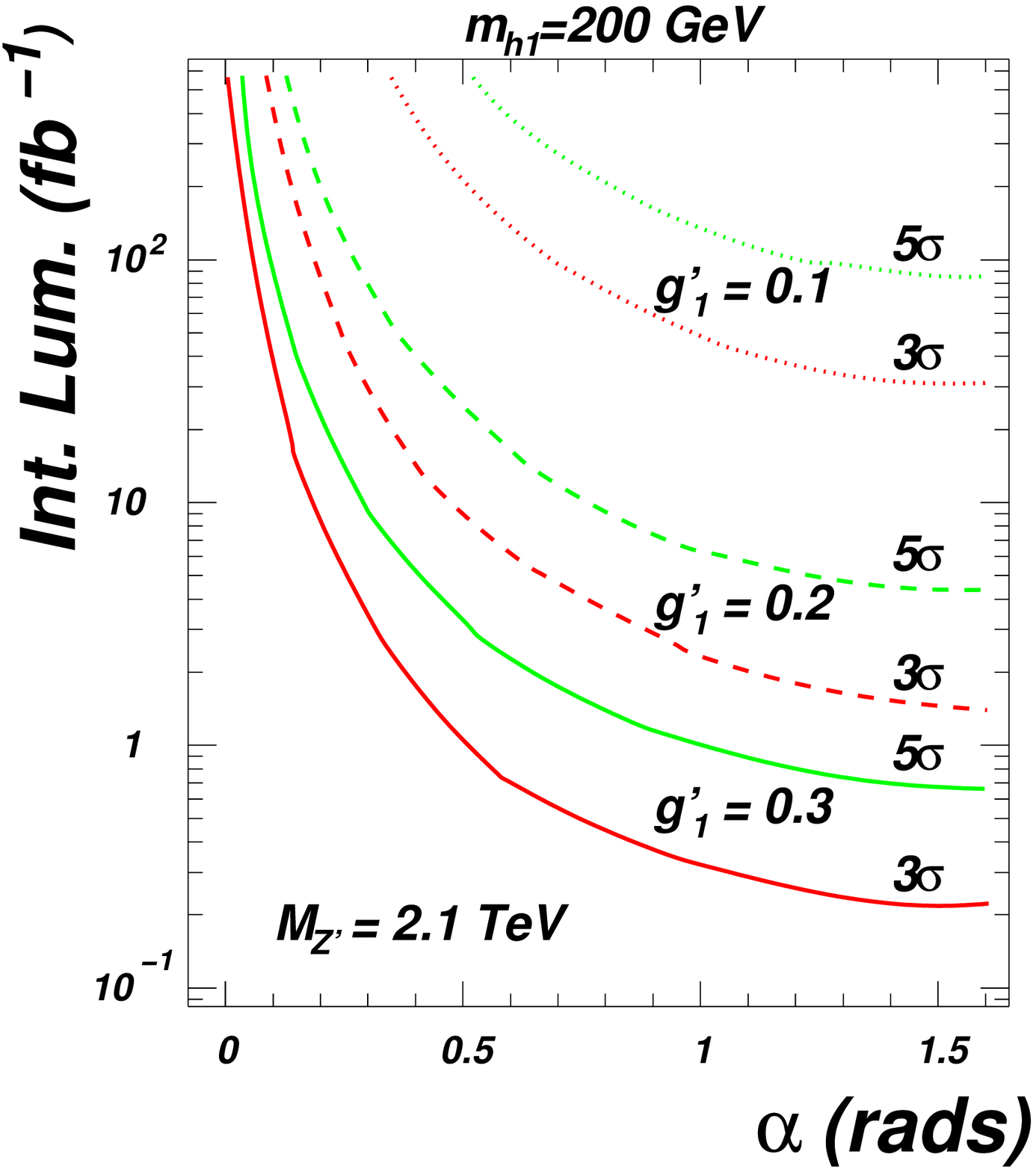}}
  \caption{\it Discovery potential for the associated production of
    the $Z'$ boson and a light Higgs boson decaying into
    (\ref{lumi120_2100})  $b$ quark pairs and into
    (\ref{lumi200_2100}) $W$ boson pairs for $M_{Z'}=2.1$ TeV and
    $g'_1=0.1$, $0.2$, $0.3$. \label{lumi_s_Zph1}}
\end{figure}

Figure~\ref{lumi_s_Zph1} shows the discovery reach of a LC in these
conditions, as a function of the scalar mixing angle
$\alpha$\footnote{The significance plots have been obtained using the
  same algorithms described in Reference~\cite{Basso:2010pe}.}.
The discovery power for the two decay modes of the light Higgs boson
are comparable.
Also, very small angles require high luminosity and big values of
$g'_1$ to be probed, excluding $\alpha=0$ for which the $h_1$ and $Z'$
bosons do not couple.
Numerical results for the $3$($5$)$\sigma$ discovery potential of
$h_1$-strahlung off $Z'_{B-L}$ are collected in Table~\ref{5sigma_ZpH1}.

The Higgs-strahlung off $Z'$ channel is the fundamental process for
producing the heavy Higgs boson for scalar mixing angles close to
decoupling.
Decoupled from the $SM$ particles, the heavy Higgs could be a very
long-lived particle (decaying into peculiar final states through heavy
neutrino pairs \cite{Basso:2008iv,Basso:2010yz}). When $\alpha >
10^{-8}\div 10^{-5}$ instead, $SM$ decay modes become dominant. In both cases, the mass of the heavy scalar can be measured from the $Z'$ recoil
mass.

\begin{table}[!hb]
\begin{center}
$m_{h_1}=120$ GeV
\begin{tabular}{|c||c|c||c|c|c|}
\hline
$\sqrt{s}=\sqrt{s_{MAX}}$ & \multicolumn{2}{|c||}{$M_{Z'}=1.5$ TeV} & \multicolumn{3}{|c|}{$M_{Z'}=2.1$ TeV} \\
\hline
$\alpha$ (rads) & $g'_1=0.1$ & $g'_1=0.2$  & $g'_1=0.1$  & $g'_1=0.2$   & $g'_1=0.3$  \\
\hline
0.2 & $>$500($>$1000) & 38(100)  & $>$500($>$1000) & 50(150) & 7(20) \\
0.5 & 120(350) & 4.5(15.0)      & 180(500) & 7(20) & 1.0(3.5) \\
1.0 & 30(90)   & 1.2(3.5)       & 45(120)  & 1.8(5.0) & 0.35(1.0)\\
\hline  
\end{tabular}
$m_{h_1}=200$ GeV
\begin{tabular}{|c||c|c||c|c|c|}
\hline
$\sqrt{s}=\sqrt{s_{MAX}}$ & \multicolumn{2}{|c||}{$M_{Z'}=1.5$ TeV} & \multicolumn{3}{|c|}{$M_{Z'}=2.1$ TeV} \\
\hline
$\alpha$ (rads) & $g'_1=0.1$ & $g'_1=0.2$  & $g'_1=0.1$  & $g'_1=0.2$   & $g'_1=0.3$  \\
\hline
0.2 & $>$500($>$1000)& 50(120)  & $>$500($>$1000) & 90(200) & 9(22) \\
0.5 & 150(420)   & 6.5(18.0)    & 200(500)  & 9(25) & 1.0(3.5) \\
1.0 & 35(100)    & 1.8(4.5)     & 45(120)  & 2.2(6.0) & 0.35(1.0)\\
\hline  
\end{tabular}
\end{center}
\vskip -0.5cm
\caption{\it Minimum integrated luminosities (in fb$^{-1}$) for a $3\sigma$($5\sigma$) discovery as a function of the  scalar mixing angle $\alpha$, for selected $Z'_{B-L}$ boson masses and $g_1'$ couplings for the light Higgs boson. All values above the given $\alpha$ are probed for the luminosity in Table. For $h_2$, all angles below $\pi/2-\alpha$ are probed with the luminosity in the Table.}
\label{5sigma_ZpH1}
\end{table}

In Table~\ref{5sigma_ZpH2_a0} we summarise the $3(5)\sigma$ discovery
potential for the heavy Higgs boson for scalar mixing angles in
proximity of the decoupling regime, i.e., $\alpha \sim 10^{-4}$ rads,
for selected values of $Z'_{B-L}$ masses and couplings. 

\begin{table}[!hb]
\begin{center}
\begin{tabular}{|c||c|c||c|c|c|}
\hline
$\sqrt{s}=\sqrt{s_{MAX}}$ & \multicolumn{2}{|c||}{$M_{Z'}=1.5$ TeV} & \multicolumn{3}{|c|}{$M_{Z'}=2.1$ TeV} \\
\hline
$m_{h_2} \mbox{ (GeV)}$ & $g'_1=0.1$ & $g'_1=0.2$  & $g'_1=0.1$  & $g'_1=0.2$   & $g'_1=0.3$  \\
\hline
120 & 20(55)  & 0.80(2.8)     & 30(90)  & 1.2(4.0) & 0.22(0.70) \\
200 & 20(60)  & 0.95(3.0)     & 30(90)  & 1.5(4.5) & 0.22(0.70)\\
500 & 0.07(0.2) & 1.5(4.0)   & 0.1(0.3)&   2.0(6.0)    &  20(65)  \\
\hline  
\end{tabular}
\end{center}
\vskip -0.5cm
\caption{\it Minimum integrated luminosities (in fb$^{-1}$) for a $3\sigma$($5\sigma$) discovery for selected $Z'_{B-L}$ boson masses and $g_1'$ couplings for the heavy Higgs boson when $0<\alpha \ll 1$ rads.}
\label{5sigma_ZpH2_a0}
\end{table}

\section{Higgs-strahlung off $Z'$: double Higgs production}

The $Z^{\prime\ast}\to Z'h_2 \to Z'h_1h_1$ process is a peculiar signature
of the $B-L$ model, not present in many other $SM$ extensions. A $h_2$
boson is radiated from the $Z'$ boson and it subsequently decays into
a light Higgs boson pair with cross sections much bigger than the
usual double Higgs-stralung process. It vanishes in the decoupling
regimes (both for $\alpha \equiv 0$ and $\pi/2$).

For $M_{Z'}=2.1$ TeV, a heavy Higgs boson with $500$ GeV mass can pair
produce the light Higgs boson with cross sections well above the fb
level up to $m_{h_1} = 200$ GeV, reaching $\mathcal{O}(10)$ fb for
small (but not negligible) values of the mixing angle (i.e., $\pi
/20<\alpha <\pi/5$). 

\begin{figure}[!th]
  \subfloat[]{
  \label{H2-H1H1Zp_1}
  \includegraphics[angle=0,width=0.45\textwidth ]{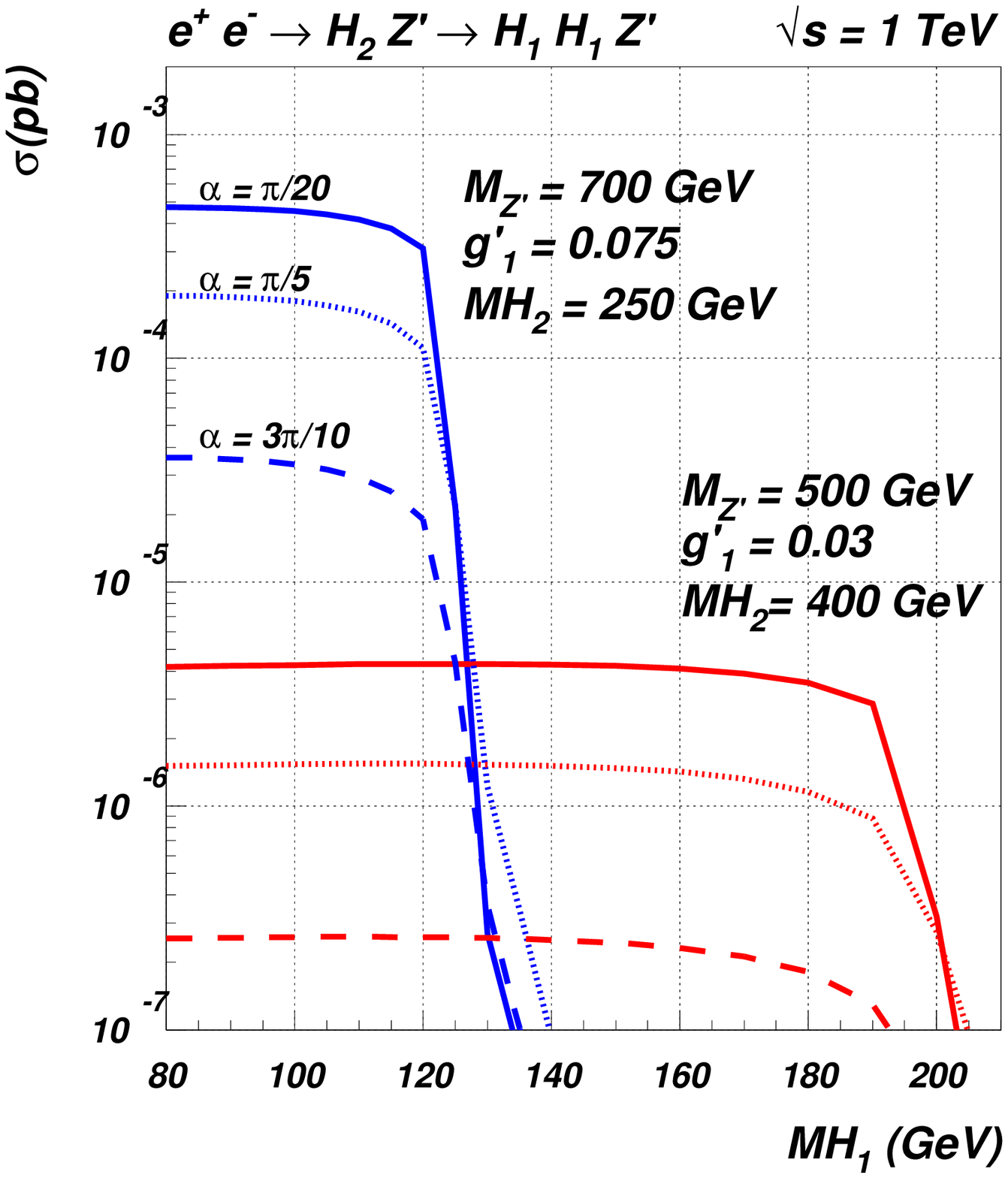}}
  \subfloat[]{
  \label{H2-H1H1Zp_3}
  \includegraphics[angle=0,width=0.45\textwidth ]{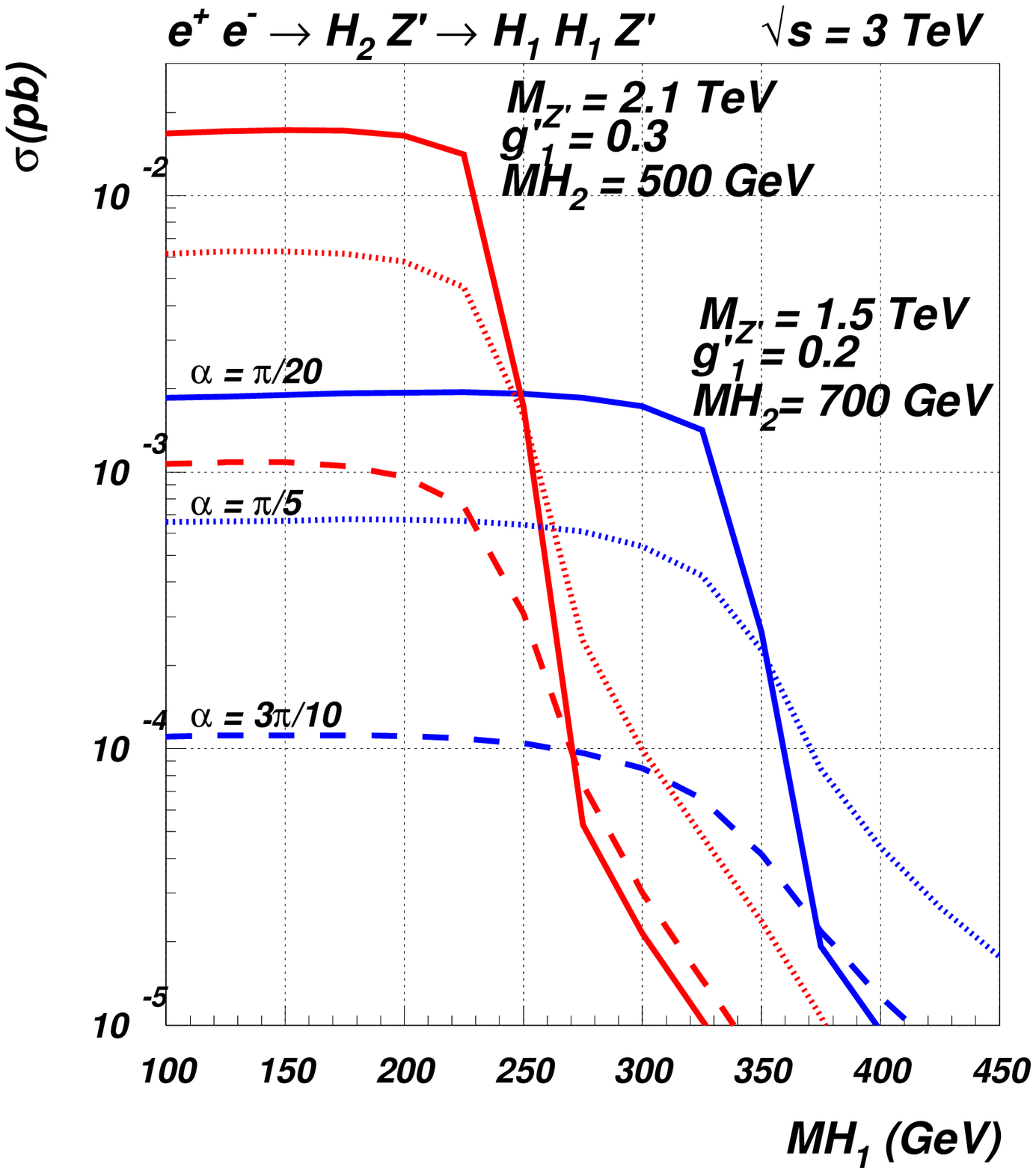}}
%  \vspace*{-0.4cm}
  \caption{\it Cross sections for the process
    $e^+e^-\rightarrow H_2 Z' \rightarrow H_1 H_1 Z'$ for $\sqrt{s}=1$
    TeV (\ref{H2-H1H1Zp_1}) and $\sqrt{s}=3$ TeV (\ref{H2-H1H1Zp_3}). \label{H2-H1H1Zp}}
\end{figure}

If a $Z'$ boson of $1.5$ TeV mass is considered, a heavier $h_1$ can be
pair produced. Cross sections are bigger than $0.1$ fb for
$m_{h_1}<350$ GeV and $\mathcal{O}(1)$ fb, again for small (but not
negligible) values of the mixing angle (see
Figure~\ref{H2-H1H1Zp}). 

\section{Other Higgs boson production mechanisms via a $Z'$
  boson}\label{sect:NS_single_other}

Finally, a possibility is to use the heavy
neutrino as a source of light Higgs bosons. It is a very peculiar
feature of the $B-L$ model, allowing for a direct measurement of the
$h_1\nu_h\nu_l$ coupling assuming that $h_1\rightarrow
\nu_h\nu_h$ is kinematically forbidden
($\nu_{h(l)}$ being the heavy(light) physical neutrino). A LC
 is again the most suitable
environment to test this mechanism, not only because of substantial
$Z'_{B-L}$ production, but also because the possibility of
tuning the $CM$ energy on the $Z'$ boson peak will
enhance the $Z'$ production cross section (and consequently the
signal) by a factor of 
roughly $10^3$. Moreover, the Branching Ratio ($BR$) of a heavy neutrino
into a light Higgs boson and a light neutrino is $\sim 20\%$ (at the
most \cite{Basso:2008iv}), when allowed. 

At an illustrative $CM$ energy of $\sqrt{s}=1$ TeV, the cross section
for $e^+e^-\rightarrow Z' \rightarrow \nu_h\nu_h\rightarrow
h_1\nu_l\nu_h$ with $M_{Z'}=700$ GeV is $\mathcal{O}(1\div 10)$~fb for a heavy neutrino of
$200$ GeV mass, decreasing to $\mathcal{O}(1)$ fb when a mass of $300$
GeV is considered, for a good range in the mixing angle. 

As intimated, on the $Z'$ peak, the $\nu_h$ pair production is
enhanced by a factor $\sim 10^3$, with cross sections
$\sim\mathcal{O}(1)$~pb for a large portion of the allowed parameter
space, and $>\mathcal{O}(10)$~fb whatever $\alpha$ is.

\section*{Acknowledgements}

GMP is thankful to Professor Shinya Kanemura for the invitation 
to the LCWS11 and to the Universidad de Granada and the TU Dresden 
for the economical support.

\noindent
LB is supported by the Deutsche
Forschungsgemeinschaft through the Research Training Group
GRK\,1102 \textit{Physics of Hadron Accelerators}.

\noindent
GMP is supported by the German Research Foundation DFG through 
Grant No.\ STO876/2-1 and by BMBF Grant No.\ 05H09ODE.

\noindent
The authors acknowledge partial financial support through the NExT Institute.

%\section{Bibliography}

%If possible please use BibTeX information as given by INSPIRE
%to make the citations~\cite{parton_qed} uniform and follow the 
%examples~\cite{parton_qed,H1,DVCS,pomeron} given below.
%Note that there is a (non-breaking) space before \verb?\cite?.

% ****************************************************************************
% BIBLIOGRAPHY AREA
% ****************************************************************************

\begin{footnotesize}
% IF YOU DO NOT USE BIBTEX, USE THE FOLLOWING SAMPLE SCHEME FOR THE REFERENCES
% ----------------------------------------------------------------------------

% ----------------------------------------------------------------------------

\end{footnotesize}

% ****************************************************************************
% END OF BIBLIOGRAPHY AREA
% ****************************************************************************


\begin{thebibliography}{99}
% Please replace the numbers for   contribId   and   sessionId
% in the following URL. You can get this information by going to 
% http://indico.cern.ch/confAuthorIndex.py?confId=2628
% and search for your contribution and click on the title
% Be aware: '&amp;' must be replaced by simple '&' as in example below
%------- replace following references ;-)
%\cite{Basso:2008iv}
\bibitem{Basso:2008iv}
  L.~Basso, A.~Belyaev, S.~Moretti and C.~H.~Shepherd-Themistocleous,
  %``Phenomenology of the minimal B-L extension of the Standard model: Z' and neutrinos,''
  Phys.\ Rev.\ D {\bf 80} (2009) 055030
  [arXiv:0812.4313 [hep-ph]].
  %%CITATION = ARXIV:0812.4313;%%
%\cite{Basso:2010yz}

%\cite{Basso:2010jt}
\bibitem{par:spa}
  G.~Cacciapaglia, C.~Csaki, G.~Marandella and A.~Strumia,
  %``The Minimal Set of Electroweak Precision Parameters,''
  Phys.\ Rev.\ D {\bf 74} (2006) 033011
  [hep-ph/0604111]; \\
  %%CITATION = HEP-PH/0604111;%%
  L.~Basso, A.~Belyaev, S.~Moretti and G.~M.~Pruna,
  %``Tree Level Unitarity Bounds for the Minimal B-L Model,''
  Phys.\ Rev.\ D {\bf 81} (2010) 095018
  [arXiv:1002.1939 [hep-ph]]; \\
%  
  L.~Basso, S.~Moretti and G.~M.~Pruna,
  %``A Renormalisation Group Equation Study of the Scalar Sector of the Minimal B-L Extension of the Standard Model,''
  Phys.\ Rev.\ D {\bf 82} (2010) 055018
  [arXiv:1004.3039 [hep-ph]]; \\
  %%CITATION = ARXIV:1004.3039;%%
%
  L.~Basso, S.~Moretti and G.~M.~Pruna,
  %``Constraining the $g'_1$ coupling in the minimal $B-L$ Model,''
  J.\ Phys.\ G G {\bf 39} (2012) 025004
  [arXiv:1009.4164 [hep-ph]].
  %%CITATION = ARXIV:1009.4164;%%
%
\bibitem{Basso:2010yz}
  %%CITATION = ARXIV:0812.4313;%%
  L.~Basso, S.~Moretti and G.~M.~Pruna,
  %``Phenomenology of the minimal $B-L$ extension of the Standard Model: the Higgs sector,''
  Phys.\ Rev.\ D {\bf 83} (2011) 055014
  [arXiv:1011.2612 [hep-ph]].
  %%CITATION = ARXIV:1011.2612;%%
%\cite{Basso:2009hf}
\bibitem{Basso:2009hf}
  L.~Basso, A.~Belyaev, S.~Moretti and G.~M.~Pruna,
  %``Probing the Z-prime sector of the minimal B-L model at future Linear Colliders in the e+ e- ---> mu+ mu- process,''
  JHEP {\bf 0910} (2009) 006
  [arXiv:0903.4777 [hep-ph]].
  %%CITATION = ARXIV:0903.4777;%%
%\cite{Assmann:2000hg}
\bibitem{Basso:2010pe}
  L.~Basso, A.~Belyaev, S.~Moretti, G.~M.~Pruna and C.~H.~Shepherd-Themistocleous,
  %``$Z'$ discovery potential at the LHC in the minimal $B-L$ extension of the Standard Model,''
  Eur.\ Phys.\ J.\ C {\bf 71} (2011) 1613
  [arXiv:1002.3586 [hep-ph]].
  %%CITATION = ARXIV:1002.3586;%%
%\cite{Tian:2010np}
\bibitem{Assmann:2000hg}
  R.~W.~Assmann, F.~Becker, R.~Bossart, H.~Burkhardt, H.~-H.~Braun, G.~Carron, W.~Coosemans and R.~Corsini {\it et al.},
  %``A 3-TeV e+ e- linear collider based on CLIC technology,''
  CERN-2000-008.
  %%CITATION = CERN-2000-008;%%
%\cite{Abe:2001pea}
\bibitem{Abe:2001pea}
  T.~Abe,
  %``A Study of topological vertexing for heavy quark tagging,''
  hep-ex/0102022.
  %%CITATION = HEP-EX/0102022;%%
%\cite{Basso:2010pe}
\bibitem{Tian:2010np}
  J.~Tian, K.~Fujii and Y.~Gao,
  %``Study of Higgs Self-coupling at ILC,''
  arXiv:1008.0921 [hep-ex].
  %%CITATION = ARXIV:1008.0921;%%
\end{thebibliography}
\end{document}